\documentclass[%
 reprint,
 amsmath,amssymb,
 aps,
pra,
]{revtex4-2}

\usepackage{graphicx}
\usepackage{dcolumn}
\usepackage{bm}

\usepackage{amsmath}
\usepackage{verbatim}     
\usepackage[normalem]{ulem}   
\usepackage{bbold}
\usepackage{graphicx}
\usepackage{dcolumn}
\usepackage{bm}
\newcommand{\braket}[2]{\langle#1|#2\rangle}  		
\newcommand{\ket}[1]{|#1\rangle}                      		
\newcommand{\bra}[1]{\langle #1|}                     		
\newcommand{\abs}[1]{\left| #1 \right|}			
\newcommand{\beq}{\begin{equation}}
\newcommand{\eeq}{\end{equation}}
\newcommand{\bei}{\begin{itemize}}			
	\newcommand{\eei}{\end{itemize}}			

\begin{document}

\preprint{APS/123-QED}

\title{Full-mode Characterisation of Correlated Photon Pairs Generated in Spontaneous Downconversion }

\author{Alessio D'Errico$^{1,*}$}
\author{Felix Hufnagel$^{1}$}
\author{Filippo Miatto$^{1,\dagger}$}
\author{Mohammadreza Rezaee$^{1}$}
\author{Ebrahim Karimi$^{1,*}$}

\affiliation{$^1$Physics Department, University of Ottawa, Advanced Research Complex, 25 Templeton, Ottawa ON Canada, K1N 6N5}

\email{Corresponding authors: aderrico@uottawa.ca \text{and} ekarimi@uottawa.ca}
\affiliation{$^{\dagger}$ Current affiliation: Xanadu, 777 Bay St. M5G2C8 Toronto, Canada }

\date{\today}




\begin{abstract}
Spontaneous parametric downconversion is the primary source to generate entangled photon pairs in quantum photonics laboratories. Depending on the experimental design, the generated photon pairs can be correlated in the frequency spectrum, polarisation, position-momentum, and spatial modes. Exploring the spatial modes’ correlation has hitherto been limited to the polar coordinates’ azimuthal angle, and a few attempts to study Walsh mode’s radial states. Here, we study the full-mode correlation, on a Laguerre-Gauss basis, between photon pairs generated in a type-I crystal. Furthermore, we explore the effect of a structured pump beam possessing different spatial modes onto bi-photon spatial correlation. Finally, we use the capability to project over arbitrary spatial mode superpositions to perform the bi-photon state’s full quantum tomography in a 16-dimensional subspace.
\end{abstract}

\maketitle

Photon pair correlations in Spontaneous parametric downconversion (SPDC) processes are ubiquitous in all the photonic degrees of freedom, thus providing a powerful tool for quantum information and computation technologies \cite{1_walborn2010spatial,2_couteau2018spontaneous}. SPDC can also be exploited to generate high-dimensional quantum states, i.e., qudits, which may be advantageous with respect to qubits in quantum information processing \cite{1_walborn2010spatial,3_mair2001entanglement,5_flamini2018photonic}. The orbital angular momentum (OAM) is among the most promising degrees of freedom for high-dimensional quantum technologies~\cite{5_flamini2018photonic,6_bouchard2018experimental}. However, there have been arguments whether photonic’s OAM is the optimal degree of freedom to increase communication capacity \cite{7_zhao2015capacity}. Most of the optics used possess the cylindrical symmetry, and therefore, the description in terms of circular beams~\cite{8_bandres2008circular}, including the so-called LG modes, provides a convenient complete basis. There has been a growing interest in exploiting single photons’ radial mode~\cite{9_karimi2014exploring,10_krenn2014generation,11_fu2018realization,12_fontaine2019laguerre,13_gu2018gouy,14_zhou2019using}, that (together with the OAM) would provide access to the full capacity for a given optical system. Experimentally, exploring the modal structure of the SPDC state has been intensely focused on its OAM content~\cite{1_walborn2010spatial,3_mair2001entanglement}. On the contrary, the radial mode decomposition is mainly considered theoretically~\cite{15_miatto2011full} with a few experimental studies \cite{4_geelen2013walsh,16_salakhutdinov2012full,17_zhang2014radial,18_zhang2018violation}. In the first attempts to investigate the LG mode radial index spectrum of SPDC experimentally \cite{4_geelen2013walsh,16_salakhutdinov2012full}, state projections were not rigorously performed on the LG basis, but only on the radial phase jump, i.e., the Walsh mode radial index \cite{4_geelen2013walsh}. Indeed, full-mode characterisation on an arbitrary basis, including the LG modes, requires precise determination of both amplitude and phase structure of spatial modes, which has recently been demonstrated for an attenuated laser beam \cite{19_bouchard2018measuring}. The filtering effect of single-mode fibres was shown to alter the detected correlations \cite{17_zhang2014radial}. In an attempt to reconstruct radial mode correlations generated by a Gaussian pump \cite{18_zhang2018violation}, the detection holograms employed in the projection would perform poorly in tomographic measurements \cite{20_bolduc2013exact}. In this Letter, we surpass the above challenges and perform the rigorous measurement of radial and OAM states, i.e., full-mode, correlations hidden in the SPDC generated from a type-I nonlinear crystal, analysing the results for different pump modes, and characterising the bi-photon correlations using full quantum state tomography in a 16-dimensional Hilbert space.

\begin{figure*}[t]
\centering
{\includegraphics[width=14.5 cm]{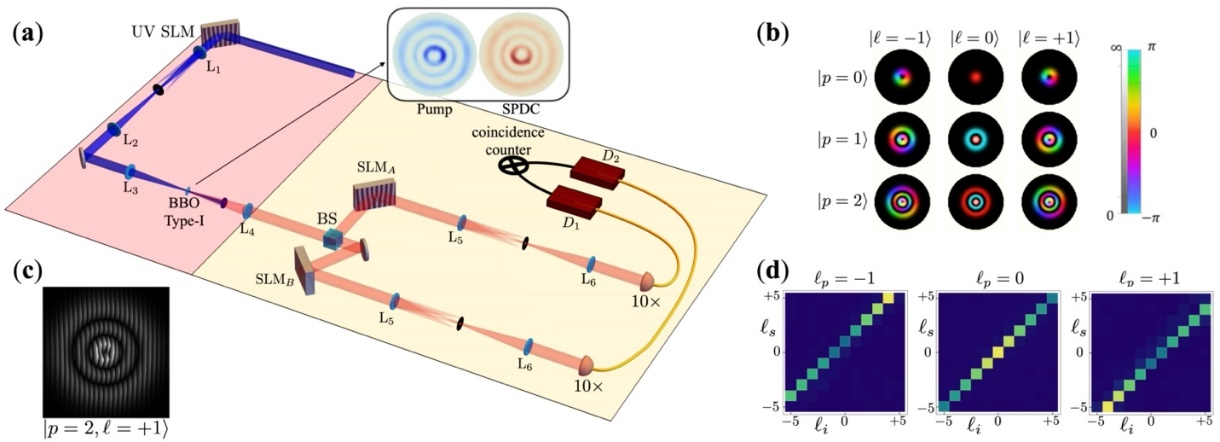}}
\caption{\textbf{Experimental setup and OAM correlations}. (a) Schematic of the experimental setup. A 400 nm laser beam converted in an LG mode exploiting an ultraviolet (UV) spatial light modulator (SLM) implementing an intensity masking technique. The beam is then focused on a type-I beta barium borate (BBO) crystal and the residual transmitted UV is filtered with a long-pass filter. The SPDC signal is collimated by a lens (L$_4$) and then sent to the detection stage through a beam splitter (BS). The individual photons are thus projected on the desired spatial modes by means of SLMs A and B and single mode optical fibres through the mode detection technique described in the text. L: lenses; Ph: pinhole; D$_1$ and D$_2$: couplers (with 10X objective) to single mode fibers and detectors. (b) Amplitude and phase distribution of the lowest order LG modes with $p\in\{0,1,2\}$ and $\ell\in\{-1,0,+1\}$. (c) Example of hologram applying intensity masking as displayed on the SLMs. The intensity masking effect can be understood by noticing that a blazed grating appears only in those regions where we wish to have a nonzero intensity. Indeed, the region with constant phase (black) will not deflect the light through the pinholes. The hologram generates LG$_{2,+1}(x)$ mode. (d) Experimental OAM correlation matrices (normalized w.r.t the maximum) for different OAM values of the pump $\ell_p$. These results have been obtained without applying intensity masking, hence both the pump and the projected modes are described as Hypergeometric-Gaussian modes.}
\label{fig:fig1}
\end{figure*}
\begin{figure}[t]
\centering
{\includegraphics[width=\linewidth]{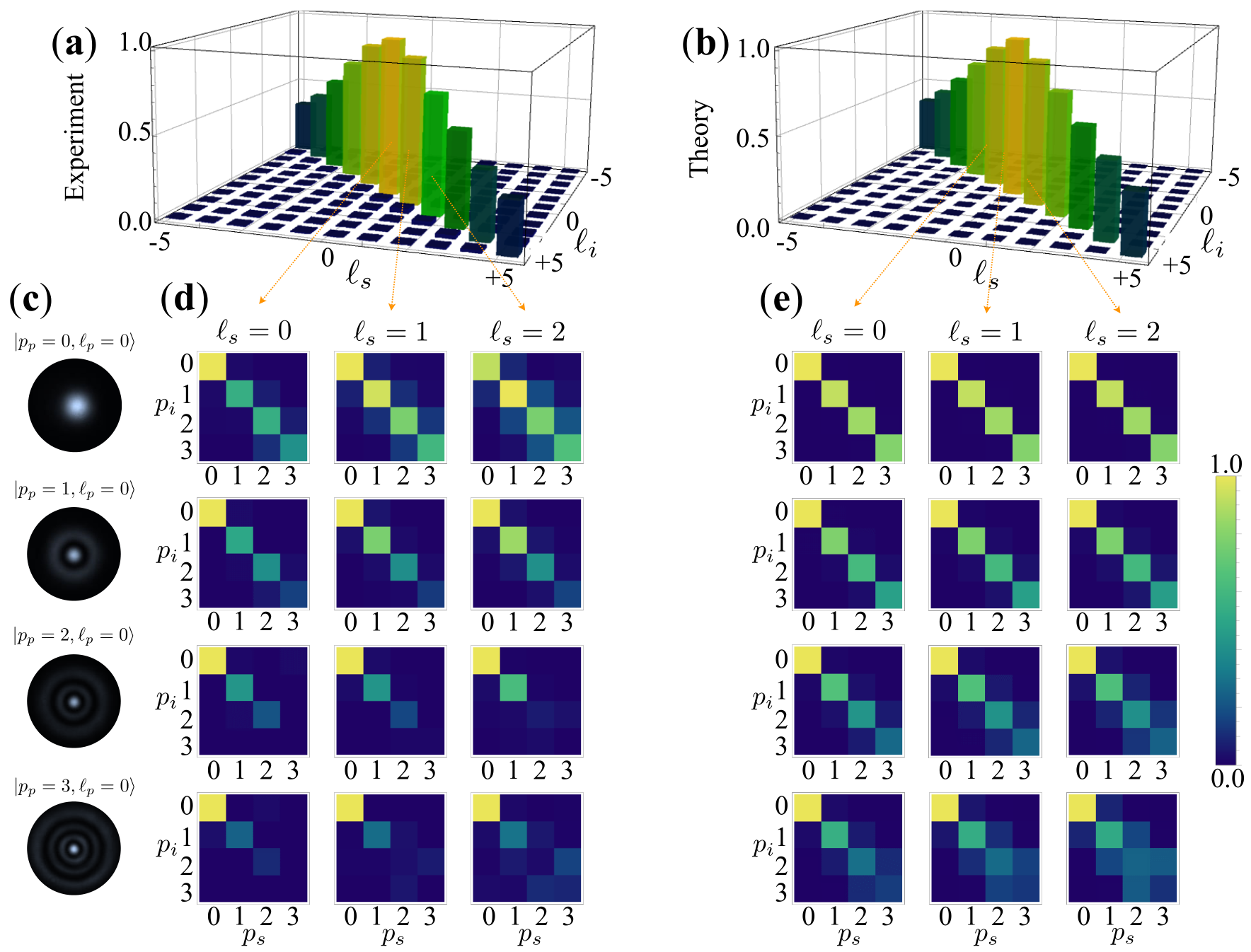}}
\caption{\textbf{ Radial mode correlations with LG$_{p,0}$ pump beam}. (a) and (b) show, respectively, experimental and theoretical OAM correlations for the case of a LG$_{0,0}$ pump. (c) Experimental pump beams intensities on the crystal plane. For each beam we show, along the same row, experimental (d) and theoretical (e) p-mode correlations of the SPDC beam. Different columns correspond to different OAM subspaces, uniquely identified by the signal OAM index $\ell_s$. For the lowest order modes (and $\ell_s$ values) we observe strong diagonal correlations.
}
\label{fig:fig2}
\end{figure}
\begin{figure}[t]
\centering
{\includegraphics[width=\linewidth]{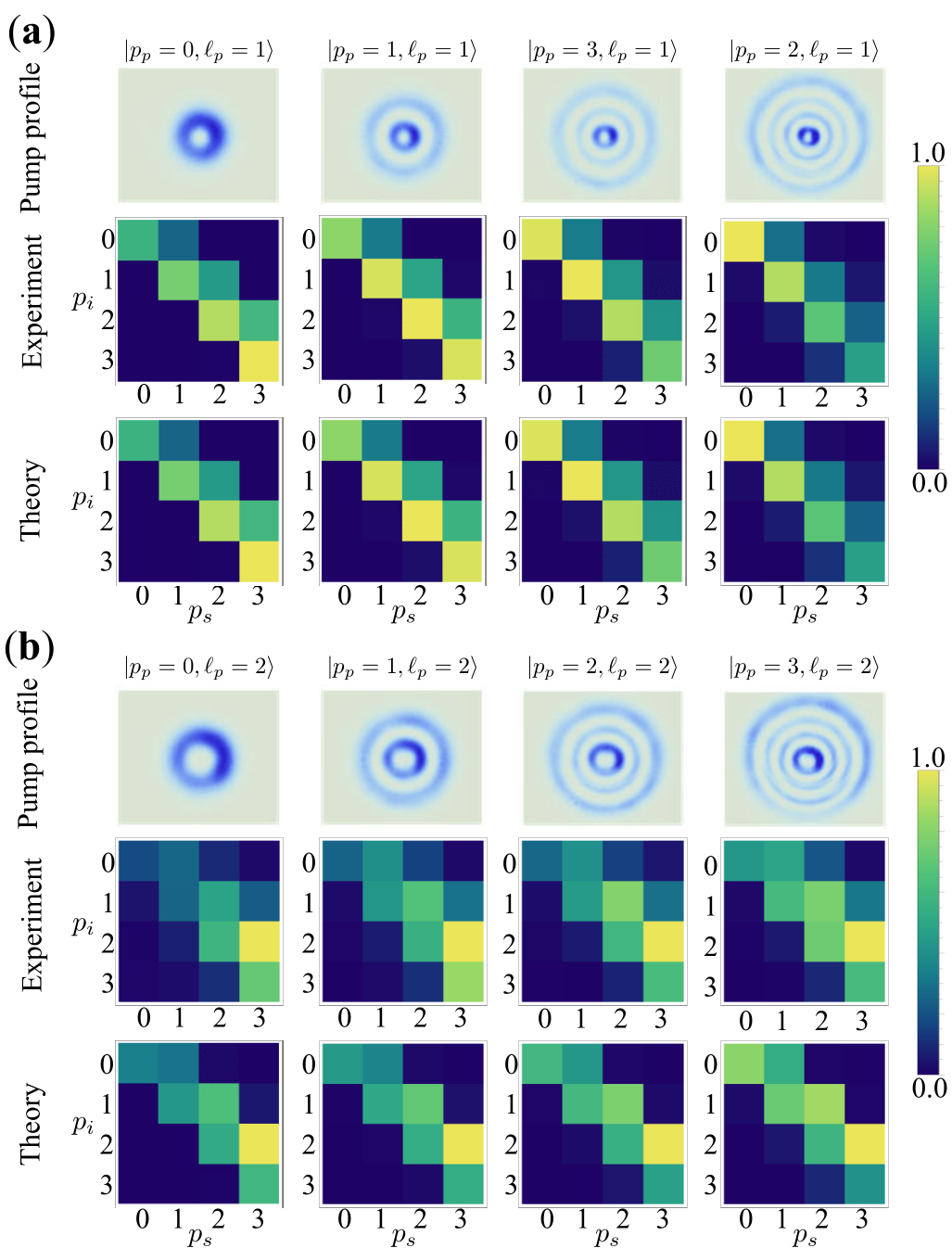}}
\caption{\textbf{Radial mode correlations with LG$_{p,\ell}$ pump beam}. (a) and (b) show experimental and theoretical radial mode correlations for pump LG$_{p,\ell}$  modes for $\ell_p=1$  and $\ell_p=2$. The corresponding intensities of the pump, measured on the crystal plane, are shown above the plots. In (a), we show the correlations for the subspace $\ell_s=2$ and $\ell_p=-1$, while in (b) the results correspond to the subspace $\ell_s=4$ and $\ell_p=-2$ .}
\label{fig:fig3}
\end{figure}
%
\begin{figure*}[t]
\centering
{\includegraphics[width=11 cm]{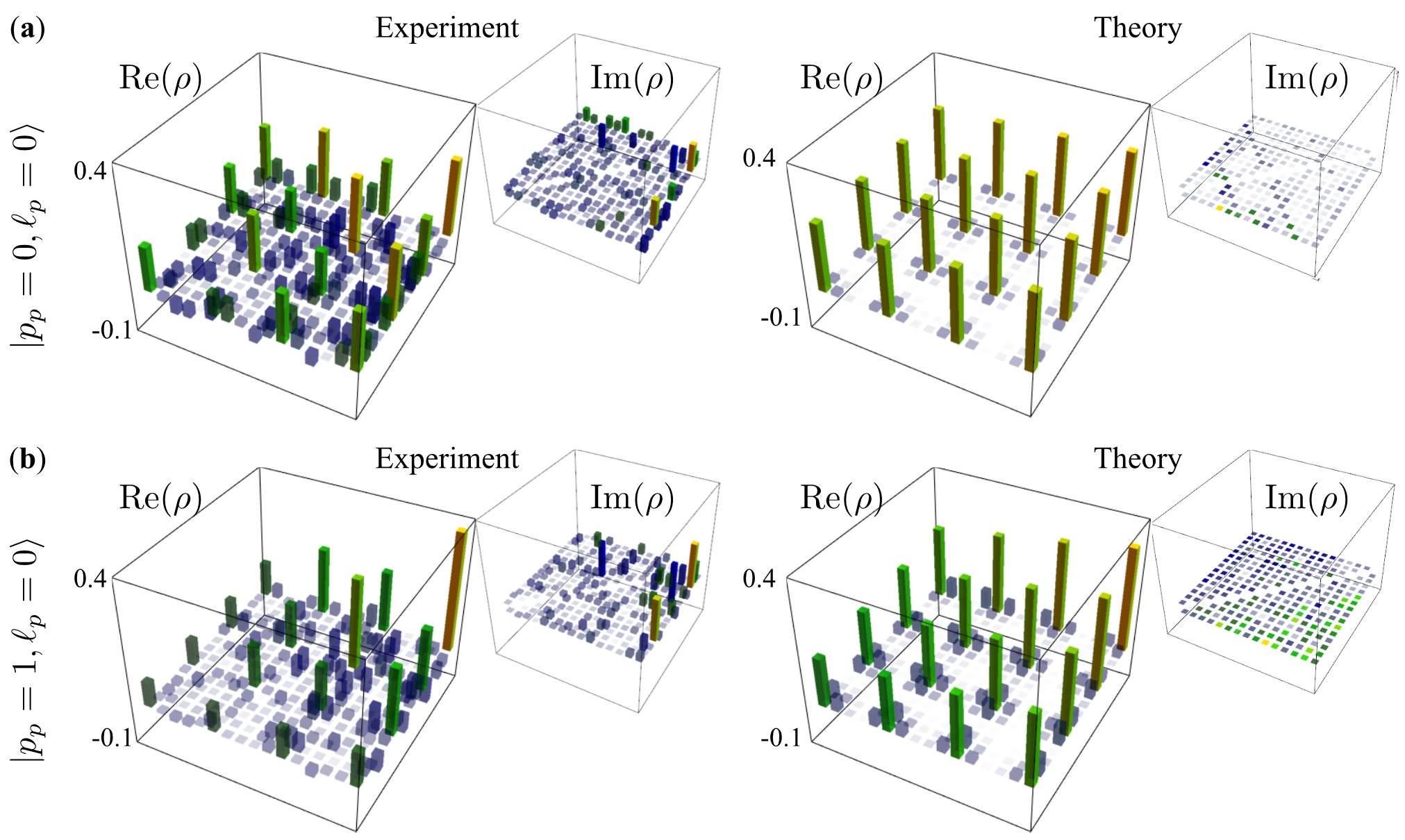}}
\caption{\textbf{State tomography for the OAM subspace}. Experimental and theoretical plots of the biphoton density matrix in an OAM subspace ($\ell_i=-\ell_s=1$) for the LG pump beam with $\ell_p=0$ and (a) $p_p=1$, and (b) $p_p=0$. We considered the subspace spanned by values of $p_s$ and $p_i$ going from 0 to 3.}
\label{fig:fig4}
\end{figure*}
In cylindrical coordinates $r,\phi,z$ one can define the complete set of LG modes labelled by two indices $\ket{p,\ell}$, determining respectively, the radial and azimuthal photon’s state. The state $\ket{\ell}$ is defined as the eigenstate of the OAM operator  $\hat{\ell}=-i\hbar\partial_\phi$ – where $\hbar$ is the reduced Planck constant, which is conjugate to the azimuthal operator $\hat{\phi}$, hence $ \Delta\hat{\phi}\Delta\hat{\ell}\geq1/4$ \cite{21_leach2010quantum}. Similarly, one can define an operator  $\hat{p}=-(\rho^{-1}\partial_{\rho}(\rho\partial_{\rho})-\rho^2+\rho^{-2}\partial_{\phi}^2-2i\partial_{\phi}+2)/4$  that is diagonal in the set of $\ket{p}$ states. However, this observable does not generate any continuous symmetry, i.e., it prevents one from finding a proper conjugate quantity $\hat{\Xi}:=\Hat{\Xi}(\hat{\rho})$ \cite{22_karimi2012radial,23_karimi2014radial}. Nevertheless, the quantum nature of $\ket{p}$ states is well-established ~\cite{9_karimi2014exploring,10_krenn2014generation,11_fu2018realization,12_fontaine2019laguerre,13_gu2018gouy,14_zhou2019using}; moreover, an uncertainty relation still holds $\Delta\hat{p}\Delta\hat{\Xi}\geq1/4$ and quantum states saturating the uncertainty relation can be engineered \cite{23_karimi2014radial,24_plick2015physical}. The explicit expression for LG modes in the position representation LG$_{(p,\ell)} (r,\phi,z):=\braket{r,\phi,z}{p,\ell}$, where the beam radius is minimized, i.e., at $z=0$, is given by,
\begin{equation}
   \text{LG}_{\ell,p}(r, \phi, 0)=C_{\abs{\ell}}^{(p)}\, \left(\frac{r}{w}\right)^{\abs{\ell}}L_{p}^{\abs{\ell}}\left(2\left(\frac{r}{w}\right)^2\right)\text{e}^{-(\frac{r}{w})^2-i\ell \phi}
    \label{eq:1lg_mode}
\end{equation}
where $C_{\abs{\ell}}^{(p)}$ is a constant and $L_{p}^{\abs{\ell}}(x)$ is the associated Laguerre polynomial \cite{25_siegman1986lasers}. Let us consider a type-I nonlinear crystal that is pumped by an ultraviolet laser beam whose complex amplitude is described by Eq. \ref{eq:1lg_mode}, i.e., $\text{LG}_{p_p,\ell_p } (r,\phi,0)$. Assuming a nondegenerate case, probabilistically, the crystal creates two identical photons, namely \textit{signal} (s) and \textit{idler} (i), from one of the pump photons \cite{1_walborn2010spatial,2_couteau2018spontaneous}. Following the conservation of energy and linear momentum, which dictates the correlation in position and anti-correlation in momentum space, the bi-photon state can be expressed in the spatial mode basis as \cite{1_walborn2010spatial,15_miatto2011full},
\begin{align}
   \ket{\Psi}_\text{SPDC}\propto&\sum_{\ell_i,p_i, \ell_s, p_s}c_{\ell_i, \ell_s}^{p_i,p_s}\ket{p_s, \ell_s}\otimes\ket{p_i, \ell_i},
   \label{eq:2SPDC_state}
\end{align}
where $\ket{p_s, \ell_s}$ and $\ket{p_s, \ell_s}$ are the signal and idler photons’ states in the LG basis, respectively, and $c_{p_s,\ell_s}^{p_i,\ell_i }$ is the bi-photon correlation amplitude. For a collinear phase matching, the bi-photon correlation amplitude is,
\begin{align}
 c_{p_s, \ell_s}^{p_i,\ell_i}=\int d\mathbf{x}\, \text{LG}_{\ell_p, P}(\mathbf{x})\text{LG}_{\ell_i,p_i}^*(\mathbf{x})\text{LG}_{\ell_s,p_s}^*(\mathbf{x}).
 \label{eq:3SPDC_coeffs}
\end{align}
where * stands for complex conjugate. This equation shows the effect of field continuity, i.e., that the amplitude (and phase) of the bi-photon wavefunction on the crystal plane is determined by the amplitude (and phase) of the pump, which we verified experimentally as shown in the inset of Fig.~\ref{fig:fig1}-a. An explicit expression for the bi-photon correlation amplitude can be found in terms of Lauricella’s Hypergeometric function – see Supplementary Information 1 for more details. The amplitude $|c_{p_s,\ell_s}^{p_i,\ell_i} |^2$ can be measured experimentally by implementing projection operators on Laguerre-Gauss modes, $\hat{P}_{p_s\ell_s}^{p_i\ell_i }=(\ket{p_s,\ell_s}\otimes\ket{p_i,\ell_i})(\bra{p_s,\ell_s}\otimes\bra{p_i,\ell_i})$, applied on each photon in the downconverted pair, i.e., $\text{Tr}(\hat{R}_{\Psi}\hat{P} _{p_s\ell_s}^{p_i\ell_i })$ – here $\hat{R}_{\Psi}$ and Tr(.) stand for the bi-photon density matrix and the trace, respectively. The measurement of the OAM content of a single photon is well-established \cite{3_mair2001entanglement}, and is typically based on the use of phase holograms (implementing a shift in the OAM space) coupled to single mode fibers – the phase flattening technique. However, projecting over spatial modes with an arbitrary amplitude shape has been for a long time a challenging task. Here, we adopt a recently introduced approach that allows, at the expense of losses, detection of LG modes (or any arbitrary set of paraxial beams) with arbitrary accuracy \cite{19_bouchard2018measuring} – see the Supplementary Information 2 for more details.
\\
Fig. \ref{fig:fig1} shows the sketch of the experimental setup – a more detailed setup is shown in Supplementary Information 3. A liquid crystal spatial light modulator (SLM) is used to shape a 400 nm pump into LG modes \cite{20_bolduc2013exact}. Idler and signal photons emitted by a type-I beta barium borate (BBO) crystal are analyzed by means of SLMs coupled to single mode fibres through a de-magnifying system (de-magnification factor=1/4) and 10X objectives, thus implementing the spatial mode projection technique \cite{19_bouchard2018measuring}. To take into account mode-dependent detection efficiencies – i.e., the fact that detection efficiency is not constant for all spatial modes – we performed calibration measurements for each state (see Supplementary Information 5 for more details on the calibration process). We select downconverted photons at the same frequency with 10 nm bandwidth filters centered around 800 nm in front of the fibre couplers. To check the alignment of the setup we first measured the correlations between signal and idler OAM states, i.e., $\ket{\ell_s}$  and $\ket{\ell_i}$, determined by the OAM of the pump, $\ket{\ell_p}$. Due to OAM conservation \cite{1_walborn2010spatial,3_mair2001entanglement,15_miatto2011full}, we detect coincidences only if $\ell_s+\ell_i=\ell_p$ (see Fig. \ref{fig:fig1}-(c), and Supplementary Information 4 for a discussion about the correlation shapes). After setting $\ell_s$ and $\ell_i$, we explore the correlation matrices for the $p$-index of the LG mode, i.e., we measured the quantities $\mathcal{P}_{p_s,\ell_s}^{p_i,\ell_i}=\abs{c_{p_s,\ell_s}^{p_i,\ell_i}}^2$ with $\ell_s+\ell_i=\ell_p$. The experimental results, see Figs. \ref{fig:fig2} and \ref{fig:fig3}, are compared with theoretical estimates based on Eq. \ref{eq:3SPDC_coeffs} where the LG modes of signal and idler are considered with a waist parameter that is 0.2 times the waist of the pump. This value has been chosen as the one which gives the best agreement with the experimental data. We performed the experiments for $\ell_p={0,1,2}$ varying the pump radial index $p_p$ from $p_p=0$ to $p_p=3$. Fig. \ref{fig:fig2} shows the re1sults relative to the case $\ell_p=0$ for some fixed values of signal and idler OAM subspaces. For low radial pump modes $p_p={0,1}$, we observe diagonal correlations between the radial indices of signal $p_s$ and idler photon $p_i$, with small variations in the different subspaces. In general, off-diagonal correlations become more relevant either by increasing the pump radial index or the OAM subspace. Similar results for nonzero values of the pump OAM $\ell_p={1,2}$ are shown in Fig. \ref{fig:fig3}.  In this case, we see that the different OAM absolute values of signal and idler photon are associated with an asymmetry in the correlation matrices.
Finally, we exploit our possibility to project the SPDC state onto arbitrary superposition states to perform the quantum tomography of a state defined in a 16-dimensional subspace of spatial modes. We fixed the OAM state as $\ket{\ell_i,\ell_p-\ell_i}$ and span the radial index of biphoton to $p_i,p_s={0,1,2,3}$. Such a state can be reconstructed using the procedure reported in \cite{26_langford2004measuring,27_agnew2011tomography}. The results of quantum state tomography for different pump states of $\ell_p=0$ and $p_p={0,1}$ are shown Fig. \ref{fig:fig4}. The experimental results have a fidelity with theoretical prediction $\mathcal{F}=0.71\pm0.01$ for $p_p=0$ and $\mathcal{F}=0.67\pm0.01$ for $p_p=1$. The relatively low values of the Fidelity can be ascribed to experimental issues such as dark counts as well as crosstalk effects due to low count rates, which can be reduced employing detectors with better quantum efficiency and lower dark counts. Notwithstanding, d=16 states with fidelity reported here would violate generalized Bell inequalities \cite{27_agnew2011tomography}, and thus can be employed in high-dimensional quantum information processing such as high-dimensional quantum teleportation and communication. 
\\
We conclude remarking that our analysis applies to any set of paraxial modes that can be reliably produced using a phase-only spatial light modulator, including transverse momentum modes, as recently shown in \cite{28_valencia2020high}. Our approach allows one to characterise the full spatial mode of bi-photon states, employing quantum state tomography beyond the OAM space. Introducing and employing radial modes, together with OAM, significantly increases the Hilbert space at the disposal of photonic quantum information processing without the need to reach mode orders with a large divergence.

\section{Funding} This work was supported by Canada Research Chairs (CRC), Canada First Research Excellence Fund (CFREF) Program, and NRC-uOttawa Joint Centre for Extreme Quantum Photonics (JCEP).

\section{Acknowledgments} The authors would like to thank Sajedeh Shahbazi and Florian Brandt for their first attempt to design the experimental setup.
\section{Disclosures} The authors declare no conflicts of interest.\\

\section{Data availability} Data underlying the results presented in this paper are not publicly available at this time but may be obtained from the authors upon reasonable request.

\section{Supplemental document}
See Supplement 1 for supporting content. 

\bibliography{sorsamp}

\onecolumngrid
\newpage
\vspace{1cm}
\begin{center}
    \Large \textbf{Supplement 1}
\end{center}
\vspace{0.2cm}
\section{Solution of Equation (3)}
 Eq. (3), after solving the azimuthal integration, which gives OAM conservation, can be put in the form

\begin{align}
c_{p_s, \ell_s}^{p_i,\ell_i}=&\mathcal{N}\int_0^{\infty} dr\, r^{\abs{\ell_i}+\abs{\ell_s}+\abs{\ell_p}+1}e^{-r^2(/1w_i^2+1/w_s^2+1/w_p^2)}\nonumber\\& \times L_{p_i}^{\abs{\ell_i}}(2(r/w_i)^2)L_{p_s}^{\abs{\ell_s}}(2(r/w_s)^2)L_{p_p}^{\abs{\ell_p}}(2(r/w_p)^2),
\label{eq:theory_explicit}
\end{align}
which can be solved analytically in terms of the first Lauricella hypergeometric function $F^{(3)}_A$ as we show below. In general the function $F^{(3)}_A$ has to be evaluated numerically, for example exploiting its integral representation, hence this result has no particular advantage with respect to the simple numerical evaluation of Eq. \ref{eq:theory_explicit}. However a more numerically accessible analytical formula can be given in the case $p_p=0$, using $L_{0}^{\abs{\ell_p}}(x)=1$. From the results in Ref. \cite{Lee2001} one can easily obtain:
\begin{align}
c_{p_s, \ell_s}^{p_i,\ell_i}(p_p=0,\ell_p)=\,&\mathcal{N}\binom{p_i+\abs{\ell_i}}{p_i} \binom{p_s+\abs{\ell_s}}{p_s}\frac{\Gamma (\ell_T+1)}{\sigma^{\ell_T+1}}\nonumber\\&
\times F_2[\ell_t+1,-p_i,-p_s; \abs{\ell_i}+1,\nonumber\\&\abs{\ell_s}+1; \lambda_i/\sigma,\lambda_s/\sigma]    
\end{align}
where $F_2[a,b,b',;c,c';x,y]$ is the second Appell function (which can be computationally implemented easily in MAPLE), $\ell_t=(\abs{\ell_p}+\abs{\ell_i}+\abs{\ell_s})/2$, $\sigma=(1/w_p^2+1/w_i^2+1/w_s^2)$ and $\lambda_{i,s}=2/w_{i,s}^2$. $\mathcal{N}$ is a normalization constant given below for the general case of arbitrary $p_p$.

For the general case, We start from Eq. \ref{eq:theory_explicit} which, with $r^2\equiv x$, can be rewritten as:
\begin{align}
    c_{p_s, \ell_s}^{p_i,\ell_i}=\mathcal{N}\int_0^{\infty}x^{\ell_T}e^{-\sigma x}L_{p_p}^{\abs{\ell_p}}(\lambda_p x)L_{p_i}^{\abs{\ell_i}}(\lambda_i x)L_{p_s}^{\abs{\ell_s}}(\lambda_s x)dx
    \label{eq:3LG}
\end{align}
where $\lambda_p=2/w_p^2$ and
$$ \mathcal{N}=2\pi\sqrt{\frac{2\, p_p!p_i!p_s!}{\pi^3 (p_p+\abs{\ell_p})!(p_i+\abs{\ell_i})!(p_s+\abs{\ell_s})!}}\frac{\sqrt{2}^{2\ell_T}}{w_p^{\abs{\ell_p}+1}w_i^{\abs{\ell_i}+1}w_s^{\abs{\ell_s}+1}}.$$
Eq. \ref{eq:3LG} can be immediately solved using Eq. 4 in Ref. \cite{Lee2001}. We obtain
\begin{align}
    c_{p_s, \ell_s}^{p_i,\ell_i}=&\mathcal{N}\binom{p_p+\abs{\ell_p}}{p_p}\binom{p_i+\abs{\ell_i}}{p_i}\binom{p_s+\abs{\ell_s}}{p_s}\frac{\Gamma(\ell_T+1)}{\sigma^{\ell_T+1}}\nonumber\\&\times F_A^{(3)}[\ell_T+1,-p_p,-p_i,-p_s;\abs{\ell_p}+1,\abs{\ell_i}+1,\abs{\ell_s}+1; \frac{\lambda_p}{\sigma},\frac{\lambda_i}{\sigma},\frac{\lambda_s}{\sigma}].
    \label{eq:3LGsolved}
\end{align}
$F_A^{(3)}$ is the first Lauricella's hypergeometric function, defined by the series:
\begin{align}
F_A^{(3)}[\ell_T+1,-p_p,-p_i,-p_s;&\abs{\ell_p}+1,\abs{\ell_i}+1,\abs{\ell_s}+1; \frac{\lambda_p}{\sigma},\frac{\lambda_i}{\sigma},\frac{\lambda_s}{\sigma}]:=\nonumber\\ \sum_{k_1,k_2,k_3=0}^{\infty}&\frac{(\ell_T+1)_{k_1+k_2+k_3}(-p_p)_{k_1}(-p_i)_{k_2}(-p_s)_{k_3}}{(1+\abs{\ell_p})_{k_1}(1+\abs{\ell_i})_{k_2}(1+\abs{\ell_s})_{k_3}}\nonumber\\&\times\frac{(\lambda_p/\sigma)^{k_1}(\lambda_i/\sigma)^{k_2}(\lambda_s/\sigma)^{k_3}}{k_1!k_2!k_3!}
\end{align}
where $(\alpha)_k:=\Gamma(\alpha+k)/\Gamma(\alpha)$ is the Pochhammer symbol.

\section{Intensity masking technique for generating and measuring optical modes}
An exact method for generating arbitrary paraxial beams from an input plane wave was devised in~\cite{bolduc:13}. The desired beam can be obtained by selecting the first diffraction order of a phase mask described by the function, 
\begin{equation}\label{eq:amplitude_masking}
   g(x,y)=M(x,y)\,\text{Mod}\left(\frac{2\pi}{\Lambda} x+F(x,y), 2\pi\right), 
\end{equation}
with,
\begin{align}\label{eq:amplitude_masking_functions}
  M(x,y)&=1-\frac{1}{\pi}\text{sinc}^{-1}(A(x,y))\\
  F(x,y)&=\Psi(x,y)-\pi M(x,y),
\end{align}
where $A(x,y)$ and $\Psi(x,y)$ are, respectively, the amplitude and phase of the beam that one wants to generate, i.e. $A(x,y)\,e^{i\Psi(x,y)}$.\\
Now we discuss how this technique can be used "in reverse", i.e. to measure instead of generating a desired mode. 
Assume the input field has a complex amplitude $X(\mathbf{r})$. To project on a mode $\Pi(\mathbf{r})$, the intensity masking system is set to display $\Pi^*(\mathbf{r})$. The resulting field will be $X(\mathbf{r})\Pi^*(\mathbf{r})$. When this field is focused on the single mode fiber tip, it will be given by its 2D Fourier transform: $\mathcal{F}(X \Pi^*)=\mathcal{F}(X)*\mathcal{F}(\Pi^*)$ (where $*$ is the convolution operation). The detected intensity (or count rate) is proportional to
\begin{equation}
    R\propto \abs{\int\int d^2\mathbf{r}\, [\mathcal{F}(X)*\mathcal{F}(\Pi^*)](\mathbf{r})\, e^{-\frac{r^2}{\sigma^2}}}^2,
\end{equation}
where $\sigma$ is the waist of the fiber mode (approximated with a Gaussian). In the limit $\sigma\rightarrow 0$ the count rate becomes proportional to the absolute square of the Hermitian product between the two modes: $R\propto \abs{\braket{\mathcal{F}(\Pi)}{\mathcal{F}(X)}}^2=\abs{\braket{\Pi}{X}}^2$. This condition can be achieved by making the mode of the fiber, \textit{on the phase mask plane}, much larger than the mode $\Pi$. This is experimentally achieved by a magnification system mentioned in point 2). In practice, a finite $\sigma$ will lead to cross-talk effects that have to be quantified either theoretically, or from calibration measurements. Employing the same approach on photon pairs, is straightforward to see that coincidence measurements will be proportional to the projection on biphoton states $\ket{\Pi_1}\otimes\ket{\Pi_2}$, with $\Pi_{1,2}$ arbitrary spatial modes. Indeed we can show that the coincidence count rate is:
\begin{align}
    C\propto\abs{\sum_{p_s,\ell_s,p_i,\ell_i} c_{p_s, \ell_s}^{p_i,\ell_i}\braket{\Pi_1}{\ell_i, p_i}\braket{\Pi_2}{\ell_s, p_s}}^2
    \label{eq:coinc_rate}
\end{align}
which is equal to $\abs{c_{p_s, \ell_s}^{p_i,\ell_i}}^2$ when one projects on the states $\ket{\Pi_{1,2}}=\ket{\ell_{1,2},p_{1,2}}$.

In the following we give a detailed derivation of Eq. \ref{eq:coinc_rate}. To calculate the expected coincidence rate $C$ we start from the SPDC state at the nonlinear crystal plane:
\begin{align}
   \ket{\Psi}_\text{SPDC}\propto&\int d\mathbf{x}\, \tilde{{\cal E}} (\mathbf{x})\ket{\mathbf{x}}_i\otimes\ket{\mathbf{x}}_s,
   \label{spdc_BBOplane1}
\end{align}
which can be decomposed in any orthogonal set of modes $\{\ket{\tilde{f}_{a}}\}_{a\in \mathcal{I}}$, where $\mathcal{I}$ is a set of indices ($a=(p,\ell)$ in the case of LG modes):
\begin{align}
   \ket{\Psi}_\text{SPDC}\propto&\sum_{a, b}\int d\mathbf{x}\, \tilde{{\cal E}}(\mathbf{x}) \tilde{f}_a^*(\mathbf{x})\tilde{f}_b^*(\mathbf{x})\ket{\tilde{f}_a}_i\otimes\ket{\tilde{f}_b}_s:=\sum_{a, b}c_{a,b}\ket{\tilde{f}_a}_i\otimes\ket{\tilde{f}_b}_s.
   \label{spdc_BBOplane2}
\end{align}
The effect of the propagation through the setup is described by operations on the vectors $\ket{\tilde{f}_a}_{i,s}$. Since these operation are spatial transformations of optical modes it is convenient to writhe $\ket{\Psi}_\text{SPDC}$ in the $\mathcal{L}^2\otimes\mathcal{L}^2$ space, (i.e. considering the biphoton wavefunction $\Psi_{SPDC}(\mathbf{x}_1,\mathbf{x}_2):=\braket{\mathbf{x}_1,\mathbf{x}_2}{\Psi_{SPDC}}$) where the action of free space propagation and optical elements can be explicitly written:
\begin{align}
  \Psi_{\text{SPDC}}(\mathbf{x}_i,\mathbf{x}_s)\propto&\sum_{a, b}c_{a,b}\tilde{f}_a(\mathbf{x}_i)\tilde{f}_b(\mathbf{x}_s).
   \label{spdc_BBOplane3}
\end{align}
In particular, the evolution from the crystal plane to the two SLMs planes, which are placed in the Fourier plane of the crystal, is given by the 2D Fourier transform: \begin{align}
\mathcal{F}[\tilde{f}_a(\mathbf{x}_i)\tilde{f}_b(\mathbf{x}_s)]=& \int d \mathbf{x} e^{i \mathbf{x}\cdot\mathbf{X_i}}\tilde{f}_a(\mathbf{x})\int d \mathbf{x'} e^{i \mathbf{x'}\cdot\mathbf{X_s}}\tilde{f}_b(\mathbf{x'})\nonumber\\:=&
 f_a(\mathbf{X}_i)f_b(\mathbf{X}_s)
\end{align}
where $\mathbf{X}_{i,s}$ are coordinates on the SLMs planes (with dimensional factors included). \\
The remaining part of the setup before the fiber couplers (SLMs plus filtering pinholes) implements the transformation: $f_a(\mathbf{X}_i)f_b(\mathbf{X}_s)\rightarrow g_1^*(\mathbf{X}_i)f_a(\mathbf{X}_i)g_2^*(\mathbf{X}_s)f_b(\mathbf{X}_s) $, where $g_{1,2}$ are the optical modes displayed on the SLMs A and B, respectively. The coupling with the single mode fibers, with the properly designed demagnification system, is equivalent to integrating the above transformation over the whole transverse space. In conclusion we obtain:
\begin{align}
    C\propto\abs{\sum_{a, b}c_{a,b}\int g_1^*(\mathbf{X}_i)f_a(\mathbf{X}_i) d\mathbf{X}_i\int g_2^*(\mathbf{X}_s)f_b(\mathbf{X}_s) d\mathbf{X}_s}^2
\end{align}
which, if $g_1=f_a$ and $g_2=f_b$, yields $C\propto \abs{c_{a,b}}^2$.

\section{Detailed setup}

\begin{figure}
    \centering
   \includegraphics[width=\textwidth]{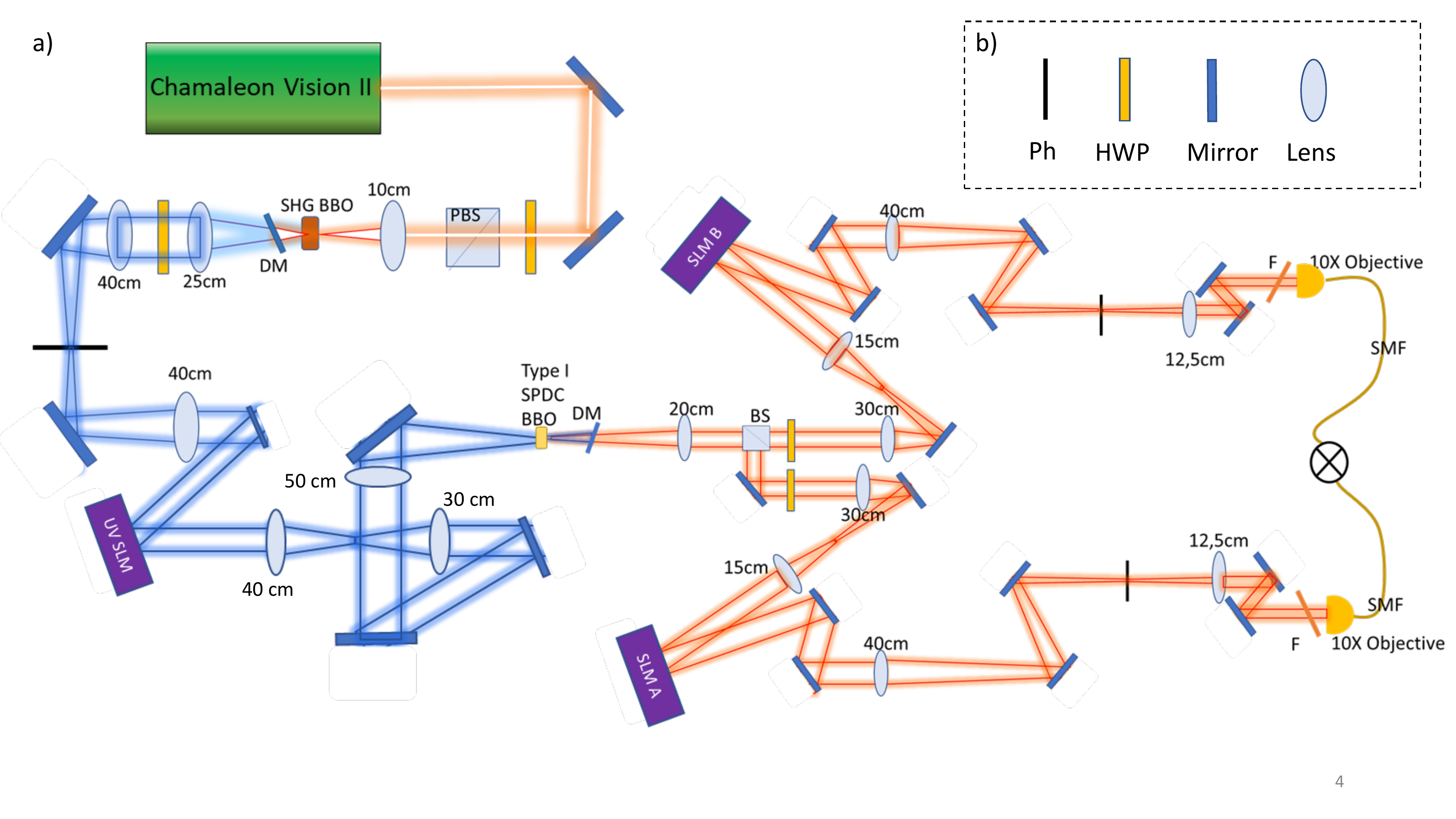}
    \caption{\textbf{Detailed experimental setup} a) Sketch of the experimental setup. b) Legend }
    \label{fig:setup2}
\end{figure}

In Fig. \ref{fig:setup2} we report a sketch of the full experimental setup. We measured a back-propagating beam waist on the SLM planes of $\sigma= 1.5$ mm, while using waist parameters on the projected modes of the order of $0.6$ mm.

\section{Simulation of OAM correlations detected without amplitude masking}

\begin{figure}
    \centering
    \includegraphics[width=\textwidth]{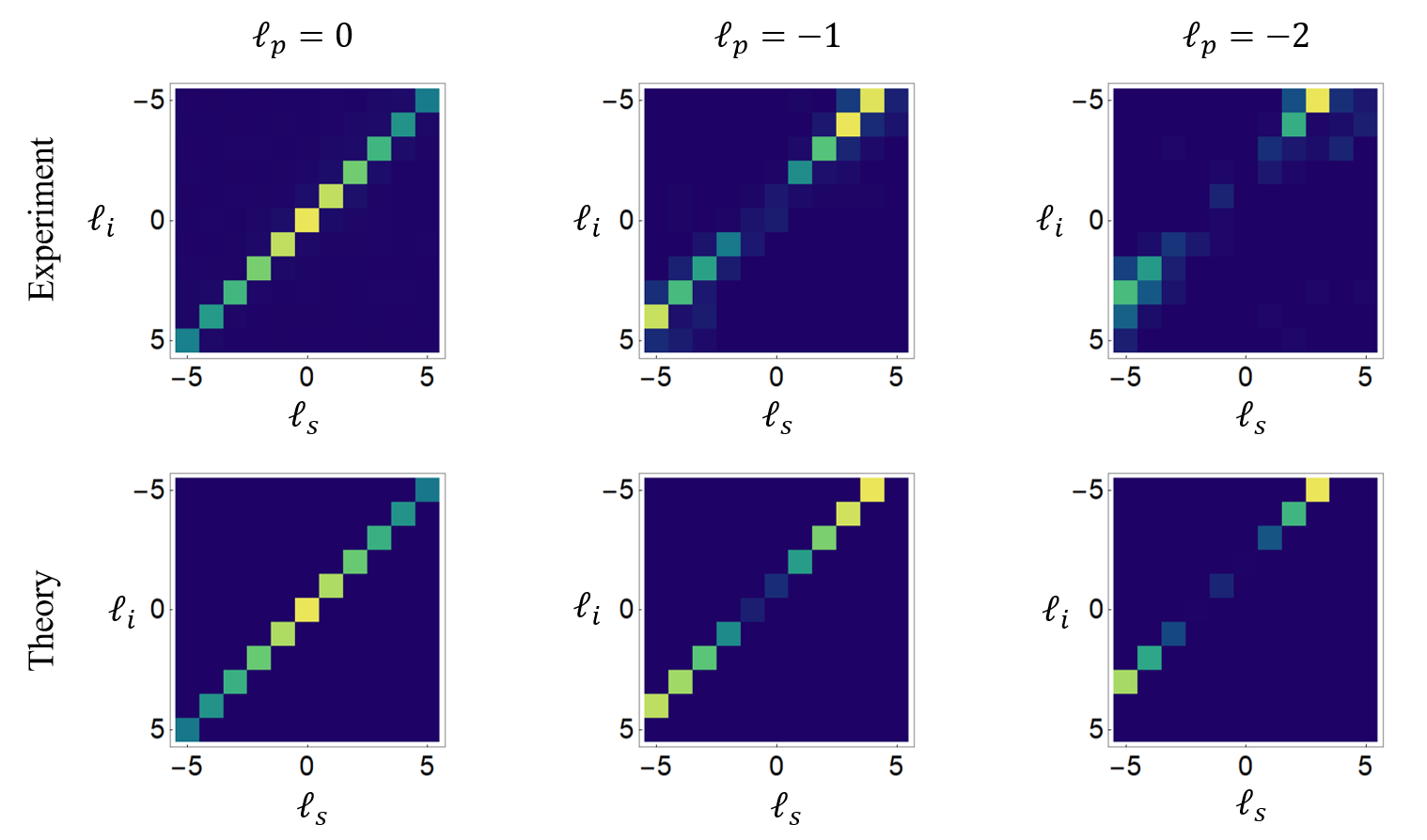}
    \caption{\textbf{OAM correlations: theory and experiment.} We compare experimentally measured OAM correlations with theoretical simulations obtained evaluating \ref{eq:corr_ell} by assuming a waist of the HyGG modes on the crystal plane $\approx 0.1 w_p$. }
    \label{fig:oamcorr_supp}
\end{figure}

 In Figures (1)-d and (2)-a,b of the main article we have shown how OAM correlations are nonzero only if $\ell_p=\ell_i +\ell_s$, i.e. one observes nonzero values along one diagonal (principal or secondary, depending on the pump OAM). It is interesting to observe the shape of the OAM correlations along these diagonals: in previous works, see e.g. \cite{bouchard:15}, one always observes that the highest coincidence rate occurs in correspondence of the lowest order modes, and this behavior is expected for any $\ell_p$. On the contrary, in our experiment we observe that, for $\ell_p\neq 0$ the correlations along the diagonals exhibit a dip in the lowest order modes. This is due to a fundamental difference between our detection system and the one used in previous works. In the latter case the detected photons where always postselected on a gaussian mode by the use of single mode fibers. In our experiment, due to the applied demagnification, we instead measure a different radial mode. If no masking is applied on the detection SLMs, then we are projecting on Hypergeometric-Gaussian modes, HyGG$_{-|\ell|,\ell}(r, z \phi)$ \cite{Karimi:07}. More specifically we may write the function $g$ displayed on the SLM as $g(r, \theta)\propto e^{-r^2}e^{i \ell_{i,s} \theta}$, where $\theta$ is the azimuthal coordinate in the SLM plane, and $r$ an adimensional radial coordinate that takes into account the finite size of the optical system (which is of the order of the backalignment beam size). The coefficients determining the coincidence rate are given by (we recall that the fields must be considered on the crystal plane):
 \begin{align}
     c_{\ell_i,\ell_s}=\int d^2\mathbf{x}\,\text{LG}_{p_p,\ell_p}(\mathbf{x})\mathcal{F}(g_{\ell_i})^*(\mathbf{x})\mathcal{F}(g_{\ell_s})^*(\mathbf{x})
     \label{eq:corr_ell}
 \end{align}
 where:
  \begin{align}
 \mathcal{F}(g_{\ell})(\mathbf{x})&=\int_0 ^{\infty} e^{-r^2} r dr \int_0^{2\pi} e^{i \ell \theta} e^{\rho r \text{sin}(\phi-\theta)} d\theta\nonumber\\
 &=(-1)^{\ell}e^{i\ell \phi}\int_0 ^{\infty} e^{-r^2} r J_{\ell}(\rho r)dr\nonumber\\
 &=(-1)^{\ell}e^{i\ell \phi}\sqrt{\pi}\frac{\rho}{8}e^{-\rho^2/8}\left(I_{(\ell-1)/2}(\rho^2/8)-I_{(\ell+1)/2}(\rho^2/8)\right),
 \label{eq:HyGG_FT}
 \end{align}
 where $I_n(z)$ are modified Bessel function of the first kind, $\rho$ the radial coordinate on the crystal plane, $\phi$ the azimuthal coordinate. Using this expression in Eq. \ref{eq:corr_ell} we obtained results in nice agreement with the observed experimental correlations (see Fig. \ref{fig:oamcorr_supp}).
 
\begin{figure}
    \centering
    \includegraphics[width=10 cm]{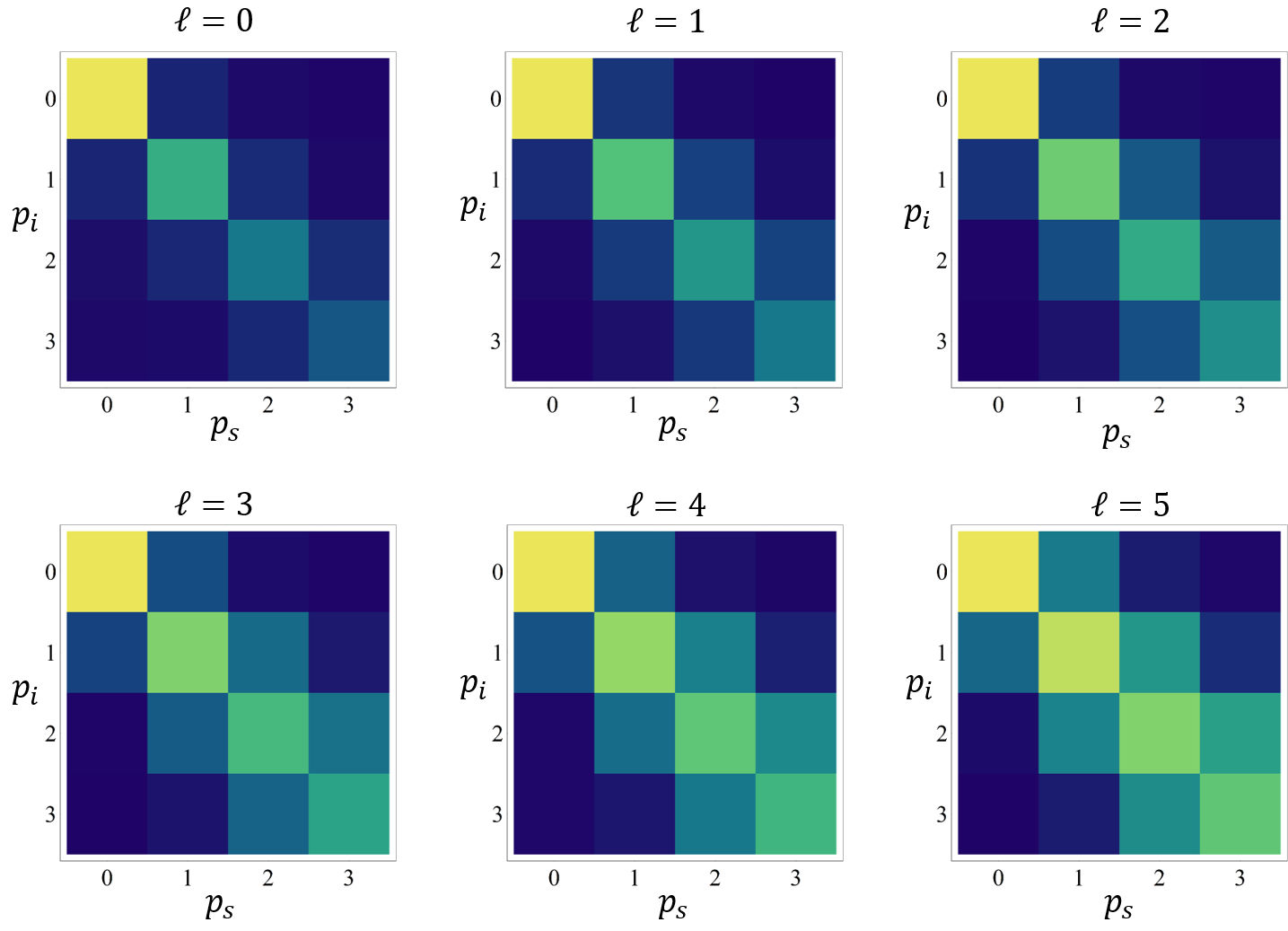}
    \caption{\textbf{detection efficiencies and cross-talk matrices} Experimentally reconstructed crosstalk matrices. The diagonal elements are proportional to the detection efficiencies used to correct the data of the main experiment. }
    \label{fig:crosstalk}
\end{figure}

\section{Measurement of cross-talk matrices and detection efficiencies}

In order to estimate the detection efficiencies we used the setup in Fig. \ref{fig:setup2} replacing the BBO crystal with a mirror and sending an 810 nm diode laser through the output coupler D$_A$. The main idea is to use SLMA to generate the desired mode and SLMB to measure it, without modifying the parameters employed in the experiment. For each generated mode (labelled as $p_i$) we measure all the radial modes $p_s=0,\ldots,3$, for a fixed OAM $\ell=0,\ldots,5$, thus retrieving the cross talk matrices in Fig. \ref{fig:crosstalk}. The (normalized) diagonal values of these matrices are proportional to the inverse of the detection efficiencies.

\section{Details on quantum tomography measurement and density matrix reconstruction}
Full quantum tomography in a $d$-dimensional Hilbert space requires projection on $d^2$ states. In our case, choosing $p_i,p_s=0\ldots,3$ we have $d=16$, hence 256 measurements are required for full tomography. The density matrix can be then reconstructed through a maximum likelihood algorithm.
 The measurement states for performing quantum tomography on a 16-dimensional Hilbert space are given by the tensor products $\ket{\psi}_i\otimes\ket{\zeta}_s$, where $\ket{\psi}$ and $\ket{\zeta}$ can be chosen among the sets: $\{\ket{p}\}_{p=0}^3$ and $\{\ket{p_1}+e^{i\alpha}\ket{p_2}\}_{p_1<p_2}$, with $\alpha=0,\pi/2$. The experimental density matrix was obtained by minimizing the quantity $\mathcal{L}(\mathbf{S}):=\sum_i [n_i-pr_i(\mathbf{S})]^2$, where the index $i$ runs over all the measured states, $n_i$ are the (normalized) count rates and $Pr_i$ the expected measurement probabilities for a target density matrix, 
\begin{align}
    \rho_\text{exp}=\sum_{i,j,l,k=0}^3 S_{i,j,k,l}\,\sigma_i\otimes \sigma_j \otimes \sigma_k \otimes \sigma_l,
\end{align} where $\{\sigma_0,\sigma_1, \sigma_2, \sigma_3\}$ are Pauli matrices (with $\sigma_0=I$) and $S_{i,j,k,l}$ are the free parameters defining the vector $\mathbf{S}$ to be found through the minimization procedure. Since the outcome of SPDC is a pure state we imposed the condition $\text{Tr}[\rho_{exp}^2]=1$.
\end{document}